# Big Data at HPC Wales

An Automated Approach to handle Data Intensive Workloads on HPC Environments

Sidharth N. Kashyap, Ade J. Fewings, Jay Davies, Ian Morris, Andrew Thomas
HPC Wales
United Kingdom
{sid.kashyap, ade.fewings, jay.davies, ian.morris, andy.thomas}@hpcwales.co.uk

Thomas Green, Martyn F. Guest
Advanced Research Computing
Cardiff University
{greent10, guestmf}@cardiff.ac.uk

*Abstract* — **This paper describes an automated approach to handling Big Data workloads on HPC systems. We describe a solution that dynamically creates a unified cluster based on YARN in an HPC Environment, without the need to configure and allocate a dedicated Hadoop cluster. The end user can choose to write the solution in any combination of supported frameworks, a solution that scales seamlessly from a few cores to thousands of cores. This coupling of environments creates a platform for applications to utilize the native HPC solutions along with the Big Data Frameworks. The user will be provided with HPC Wales APIs in multiple languages that will let them integrate this flow into their environment, thereby ensuring that the traditional means of HPC access do not become a bottleneck. We describe the behavior of the cluster creation and performance results on Terasort.**

## I. INTRODUCTION

Supercomputers represent the best in class High Performance Computing (HPC) systems and the most powerful clusters of computers available today. HPC systems have traditionally been used for Scientific Research and Discovery, environments that demand a close integration of these systems. HPC workloads have become increasingly data intensive, reflected in the associated clusters having dedicated hardware to ensure efficient I/O handling [5].

The onset of the "data deluge" problem [6] demanded solutions that could not be restricted to the niche of supercomputing alone, resulting in many companies working together to come up with Hadoop [2] - an open source platform that handles data efficiently on commodity clusters and provides a smooth and transparent scalability model.

The Hadoop and HPC workloads share common high level goals – to ensure the optimal distribution of computation across all the available compute resources while minimizing the overhead caused by I/O. HPC handles this through exclusive hardware and interconnects, whereas Hadoop relies on customized file systems and associated compute methods.

Hadoop ensures that the analysis of data is done with ease by handling the intricacies of distributed computing in its framework [2, 6]. This provides numerous opportunities for traditional HPC workloads that can now leverage the advances in Hadoop [14, 17] and the related frameworks to handle the massive growth in scientific data [18].

The underlying compute capability of Supercomputers, coupled with the growth in Hadoop as the preferred framework for data analysis, creates the need for an integration of the two environments to enable seamless execution of both workloads in HPC environments.

This paper describes ongoing work at HPC Wales [1] to create an HPC platform that efficiently handles Hadoop-based, data intensive workloads and seamlessly integrates the HPC components to ensure scalability and efficiency.

The solution (described in section III) creates a dynamic cluster based on YARN [6, 7] (Yet Another Resource Negotiator, also referred to as Map Reduce v2), when the user submits a job through the scheduler script or the API. The dynamic cluster must be created to ensure that the pre-requisites for Hadoop are satisfied, including daemons for job control and execution, environment variables customised to Hadoop and the file structures on all the nodes. The job is then executed on this cluster.

This solution presents a unified platform, including a variety of Big Data frameworks and extensions [Pig, Hive, RHadoop and MongoDB], that work together with YARN. This can be coupled with scientific workloads that traditionally use MPI, OpenMP and CUDA. The execution and orchestration of this platform is carried out through IBM Platform LSF (Load Share Facility [3]).

We chose Terasort [9] as the benchmark to fine tune the Distributed File System (Lustre [20]) and the YARN configuration parameters to achieve optimal performance on the above mentioned configuration. This is described below in the results section.

Section II provides a brief overview of the HPC Wales Infrastructure, section III an outline of the Architecture and the Design choices, while section IV summarizes the related work in this area. The YARN construction and configuration on HPC Wales is described in section V, and the experimental setup in section VI. Section VII highlights the results while section VIII presents the conclusion and outlines future work.

## II. HPC WALES

HPC Wales [1] is a large distributed supercomputing facility with nearly 17,000 cores spread across six campuses in Wales, based on a Hub-and-Spoke model. This heterogeneous architecture features both Intel Westmere and Sandy Bridge processors, along with nVidia GPUs, powering the compute plus a hierarchical storage system that includes DDN Lustre and Symantec File Stores.

We have partnered with Fujitsu for the computing services and use the Fujitsu SynfiniWay [4] framework to enable job submission via a web interface and high-level API, without the need to SSH into the system. We use IBM Platform LSF [3] as the scheduler for our clusters.

## III. ARCHITECTURE

Hadoop and HPC Environments are not typically complimentary, with work by SDSC [8] highlighting the architectural differences. The primary differences are captured in the table below [2, 5, 7].

| Feature | Hadoop | HPC |
|---|---|---|
| Scheduling | Custom scheduler FIFO/Capacity/Fairshare | System wide schedulers LSF/MOAB/Torque |
| Storage | Architected to work with DAS. HDFS[21][6] is the primary file system | Large NFS and Parallel file stores with very little DAS |
| Compute | Compute is moved to the nodes with stationary data, data moved across in stages | Custom to the application implementation |
| Interconnect | High speed interconnects are not mandated; architected to work well with Ethernet | Infiniband is the primary interconnect to ensure fast data movement |
| Services | Relies on services that control the execution of the flow and execution | Services are central to environment, typical applications do not expect custom daemons |

Given these architectural differences, we will have to make some design decisions to ensure that both environments integrate effectively.

The following are the design choices made in HPC Wales:

- **Scheduler Integration**: The Hadoop job is submitted just like any other to the job scheduler (IBM Platform LSF [3]), with the requisite resources allocated by the scheduler. The user is also provided with APIs in multiple languages that work with SynfiniWay, ensuring that compute resources can be tied to any end user application outside of the HPC Wales environment without submission via SSH.

- **Dynamic Cluster Configuration**: The Hadoop cluster is created dynamically after the requisite resources are allocated by the Platform LSF scheduler. We start the daemons on the first two nodes allocated, create the requisite directories and environment and run the application. The dynamic Hadoop cluster is torn down after job completion.

- **YARN**: Map Reduce is the underlying pattern used to program Hadoop. Yet Another Resource Negotiator (YARN) [7] is the latest version of Map Reduce which is container-based and relies on custom daemons. The container-based architecture provides the opportunity to execute generic commands and not constrain us to Map Reduce.

- **Lustre File System** [20]: We chose to use the Lustre Parallel file systems as the backend file system for Hadoop. The work by Fadika, Z. et.al. [11] demonstrate that the performance of Hadoop under regular workloads is comparable to HDFS, even when used with NFS. This gains importance as most of the HPC Wales compute nodes have very little local storage that cannot handle typical Big Data workloads (in the order of TB's).

The summarized flow of application execution in our environment is as described in Figure 1.

- **Step 1** (only if the application uses HPC Wales' APIs): The end user integrates the APIs into his program to provide the following functionalities: job submission, obtaining job status and job termination. The APIs are integrated with SynfiniWay.

- **Step 2** (only if the application uses HPC Wales' APIs): SynfiniWay submits the job into the scheduler based on the custom workflows.

- **Step 3** (only if the user submits the job through the traditional login to the cluster through SSH): The user submits his program into LSF which is integrated into a submission script.

- **Step 4**: The scheduler processes the job request received either through SynfiniWay or end user LSF submission. This results in the creation of dedicated resources for the job. The scheduler at this stage invokes the command-line associated with the job. The dynamic cluster configuration then kicks in, driven by a custom wrapper script that performs the Hadoop cluster creation – daemon initiation, directory structure creation and the environment setup. The user application is then submitted into this cluster.

- **Step 5:** Once the job is completed, the output directories and the associated job logs can be accessed by the user.

- **Step 6** (Only if the application uses HPC Wales APIs)**:** The data produced by the program can be accessed through the API or traditional access methods.

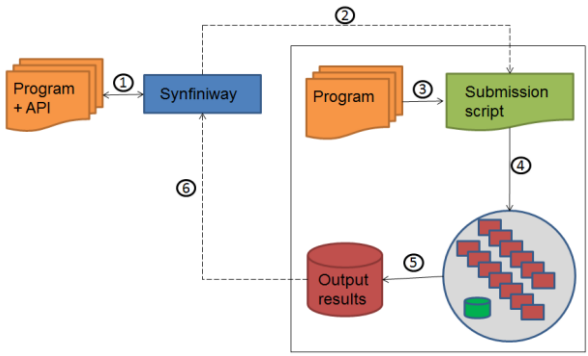

Fig. 1. Execution flow diagram

**Data Movement**: YARN expects a custom directory structure which is used by many of its daemons. It is important for us to reap the maximum benefits from the available DAS wherever possible. AC Murthy et.al. [7] provide a good overview of the daemons. We locate operational directories as follows:

- Local Directories: Application Master Log Directory, Name Node Log Directory, Resource Manager Log Directory, Name Node Data Directory.
- Lustre: Hadoop Staging, Input and Output.

## IV. RELATED WORK

Recent research by various supercomputing centers has promoted the enablement of Hadoop as the primary mode of data intensive application execution in scientific computing applications. There are over 2,000 peer-reviewed articles on such developments [10].

San Diego Supercomputer Center, one of the first to publish a comprehensive overview of using Hadoop on HPC resources, proposed myHadoop [8] as the framework for enabling Hadoop version 1 (based on Map Reduce) on regular computer clusters. Our solution follows the same philosophy as myHadoop, but is distinct in the following aspects:

- We use the latest version of Hadoop (YARN), which is not currently supported in myHadoop. This provides an opportunity to execute custom flows (anything that works as a Linux command-line works on a container).
- We use Lustre and do not configure HDFS for the reasons explained in section III.
- Our system is closely tied to Platform LSF (and SynfiniWay) which enables the custom API feature.
- We not only configure Hadoop in the environment but also enable the related frameworks such as Pig, Hive, R and Mongo DB. This provides flexibility for the application designer to use the best of all the frameworks in the solution, including native access to traditional HPC tools.

The challenge of integrating a shared HPC environment with Hadoop has been tackled by novel approaches, including the provision of a custom Map Reduce Framework that works exclusively within the HPC environment [12]. Xiaoyi Lu et.al. [15] show that the average peak bandwidth of MPICH2 is about 100 times greater than Hadoop RPC, giving us the scope to use MPI within the YARN containers. Neves et.al. [13] propose a Map Reduce adaptor that hides the HPC behavior, including the resource negotiation beforehand, as this is not the default MapReduce [19] application environment. Garza et.al. [16] suggest a solution for Hadoop on a Low-Budget General Purpose HPC Cluster - this solution proposes a pre-configured and exclusive node setup for HDFS and daemons.

There has been extensive research on enabling scientific data analysis through Hadoop [14, 17]. Manish et.al. [18] estimate the growth in scientific data that makes the successful integration of Hadoop and HPC imperative. Almeer et.al. [17] discuss the advantages of using Hadoop for Remote Sensing, while Taylor et.al. [14] overview the use of Map Reduce in Genomics.

Given the rapid development of research into provisioning and enabling Big Data flows in traditional HPC, we will need to enable the key components from these findings to ensure that scientific and traditional Big Data applications can scale and take full advantage of the HPC environment to the benefit of HPC Wales users.

## V. YARN CONSTRUCTION AND CONFIGURATION

We described the architecture of our solution in section III. As discussed, YARN brings a new computational capability to our clusters. The containers constructed and the interaction to enable job completion on our platform is described below.

The Resource Manager (RM) and per-node slave, and the Node Manager (NM) are the main components of the data-computation framework. The RM is responsible for the arbitration of resources. An Application Master Server is instantiated on one of the nodes and is responsible for the complete job execution, with the RM tracking the status of the application through the Application Master (AM). The core computational tasks are performed in the Containers instantiated on the slaves. The framework also starts the Job History Server which maintains information about MapReduce jobs after their AM terminates; this is useful in our case to debug the application [7].

We initialize the Resource Manager and Job history servers on the first two nodes allocated by LSF (as shown in Figure 2); the rest of nodes are then configured to become the slaves. This configuration is exported into the cluster environment and the daemons are triggered. Once the daemons are in place, we launch the application on the dynamically created cluster. This infrastructure is torn down after the job completes.

## VI. EXPERIMENT SETUP

The two primary aspects of investigation in our experiments are as follows:

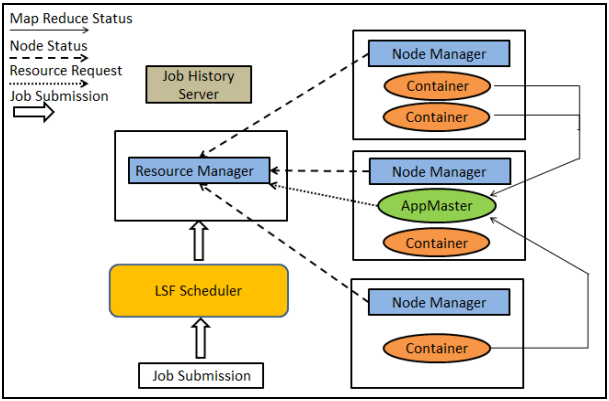

Fig. 2. YARN on HPC Wales

1. **Dynamic Cluster Setup Overhead**: We analyze the overhead of creating and tearing down the Hadoop cluster established for the user.
2. **Terasort Performance**: Terasort [9] is a widely accepted benchmark for Hadoop clusters. We used this to understand the behavior of our cluster, leading to further improvements that enhance the overall performance.

We use Lustre version 2.1.3 on Sandy Bridge machines with dual processor EP nodes (16 Cores), 414G of local storage and 64G memory per node. We submit the jobs directly into the scheduler using a separate submission node. The jobs are allocated on a dedicated queue, with exclusive access to the nodes, to ensure that performance is not impacted by shared workloads. The job script typically contains the resource request, requisite modules and HPC Wales wrapper script. The latter takes the application as an argument and is triggered after the dynamic cluster creation. The environment created contains all the Big Data frameworks supported. The infrastructure gets torn down after the application terminates.

**YARN Configuration**: Once the resource is allocated through LSF, we use the first two nodes for the daemon creation. The Key YARN parameters are as below:

| Parameter | Value |
| --- | --- |
| yarn.nodemanager.resource.memory-mb | 52GB |
| yarn.scheduler.minimum-allocation-mb | 2GB |
| yarn.scheduler.minimum-allocation-vcores | 1 core |
| yarn.app.mapreduce.am.resource.mb | 8192 |
| mapreduce.map.memory.mb | 4096 |
| mapreduce.map.java.opts | -Xmx3072m |

The number of mappers and reducers were varied in our experiments (see below).

## VII. RESULTS

We first describe the behavior of the wrapper described in step 4 of the architecture (section II). As described therein, the wrapper is responsible for the creation and the tear down of the cluster. This analysis gives us the behavioral details of the wrapper.

The results shown in Figure 3 prove that the wrapper adds little overhead to the execution. The graph compares the cores allocated to the cluster and the time taken to create the cluster. In this experiment we just create the cluster and tear it down with no time spent on the execution.

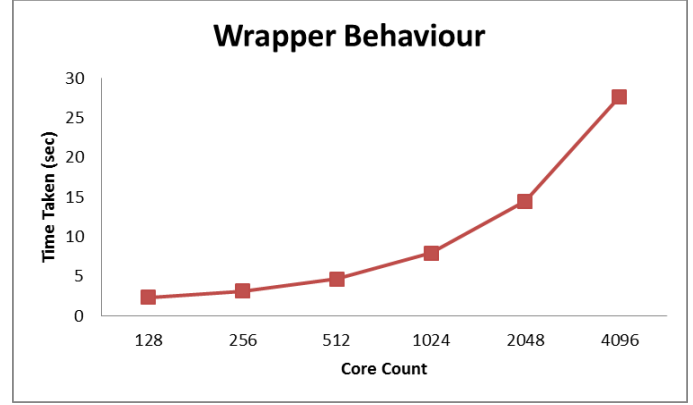

Fig. 3. Wrapper Behaviour

**Terasort Performance**: Terasort is a benchmark that tests the I/O and MapReduce components of a Hadoop cluster. This benchmark is supplied as part of the examples in the Hadoop installation. Terasort provides the opportunity to analyze the behavior of the cluster when subjected to sorting one Terabyte of data (this size can be varied). The functionality is divided into three stages, (i) Teragen, (ii) Terasort and (iii) Teravalidate. We present below the details regarding (i) and (ii) based on varied parameters.

The number of Mappers and Reducers are proportional to the allocated number of cores. The present results are based on processing a single Terabyte dataset, and are the time taken to perform the compute only.

**Teragen** [9] is a mapper only process that generates a specified size of data. In our experiments we chose to generate a Terabyte as this gives us the near worst case performance and is also the expected average workload on our systems. The Teragen performance is summarized in Figure 4. We can see that the performance is optimal at 1,800 cores, which gives us insights into the effect of the number of mappers on performance.

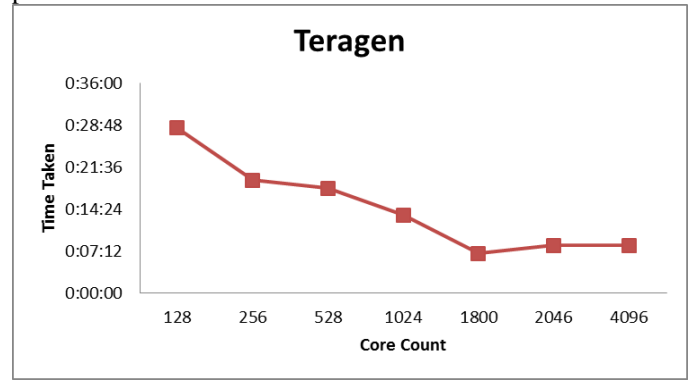

Fig. 4.  Teragen Behaviour

The Terasort program sorts the data generated by the Teragen phase. This phase of the benchmark comprises both Map and Reduce tasks, providing insight into the behavior of the cluster when the data generated in the map phase needs to be processed by the reducers. The number of mappers and reducers have been kept constant and are in line with the number of cores allocated. The Terasort performance is summarized in Figure 5.

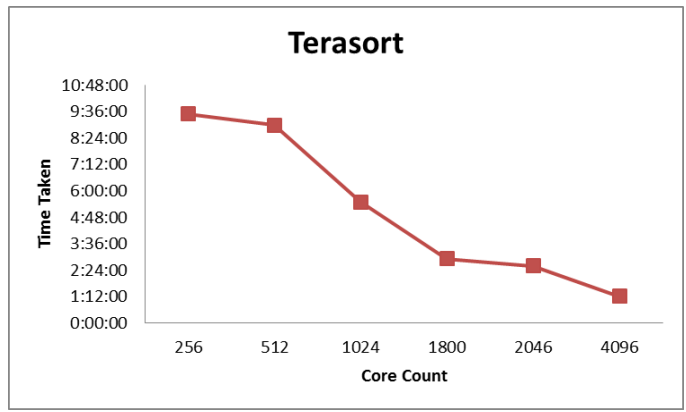

Fig. 5.  Terasort Behaviour

This dataset size is found to show reasonable scalability, but also clearly shows where we need to focus future work. Our initial investigation suggests an I/O performance bottleneck that can be influenced by the Lustre file system and Infiniband network parameters. We are continuing our efforts in this area and hope to report our findings in the near future.

## VIII. CONCLUSION AND FUTURE WORK

In this paper we have presented a description of work integrating Hadoop with an established HPC environment. We have described how we establish a Hadoop cluster dynamically as part of a job scheduled on a scheduler-controlled Supercomputer. The solution presented works on the latest version of Hadoop (2.5.1, with YARN) that presents a container-based execution model. We recognize that this is the first step in a process of integration that requires further development. The demonstrated performance is found to be modest, and provide a good base for our future planned optimization work.

This work will focus on optimising the use of the containers to vary the workloads which include non-Map reduce based jobs. We will also focus on the API's and optimizing the key components of the cluster to deliver maximum performance and reap the complete benefits of this highly scalable environment.